\input harvmac
\newcount\yearltd\yearltd=\year\advance\yearltd by 0

\noblackbox

\input epsf

\def\tilde{\widetilde}

\newcount\figno
\figno=0
\def\fig#1#2#3{
\par\begingroup\parindent=0pt\leftskip=1cm\rightskip=1cm\parindent=0pt
\baselineskip=11pt
\global\advance\figno by 1
\midinsert
\epsfxsize=#3
\centerline{\epsfbox{#2}}
\vskip 12pt
{\bf Fig.\ \the\figno: } #1\par
\endinsert\endgroup\par
}
\def\figlabel#1{\xdef#1{\the\figno}}
\def\encadremath#1{\vbox{\hrule\hbox{\vrule\kern8pt\vbox{\kern8pt
\hbox{$\displaystyle #1$}\kern8pt}
\kern8pt\vrule}\hrule}}

\def\half{{\textstyle{1\over2}}}

\def\inbar{\vrule height1.5ex width.4pt depth0pt}
\def\half{{1\over 2}}
 
 \def\m{{\mu}}
 \def\n{{\nu}}

 \def\frac#1#2{{#1\over #2}}
 \def\l{{\lambda}}

 \def\b{{\beta}}

 \def\o{{\rm ord}}

 \def\p{\partial}

 \def\r{\rightarrow}

\def\bb{{\bar{\beta}}}

\def\r{{\rho}}

\def\IR{\relax{\rm I\kern-.18em R}}
\def\IC{\relax\hbox{$\inbar\kern-.3em{\rm C}$}}
\def\IZ{\relax\ifmmode\hbox{Z\kern-.4em Z}\else{Z\kern-.4em Z}\fi}

\lref\andrei{
  A.~Mikhailov,
  ``Giant gravitons from holomorphic surfaces,''
  JHEP {\bf 0011}, 027 (2000)
  [arXiv:hep-th/0010206].
}

\lref\SeibergAX{
  N.~Seiberg,
  ``Notes on theories with 16 supercharges,''
  Nucl.\ Phys.\ Proc.\ Suppl.\  {\bf 67}, 158 (1998)
  [arXiv:hep-th/9705117].
}

\lref\BeasleyXV{
  C.~E.~Beasley,
  ``BPS branes from baryons,''
  JHEP {\bf 0211}, 015 (2002)
  [arXiv:hep-th/0207125].
}

\lref\our{
  I.~Biswas, D.~Gaiotto, S.~Lahiri and S.~Minwalla,
  ``Supersymmetric states of N = 4 Yang-Mills from giant gravitons,''
  arXiv:hep-th/0606087.
}

\lref\BerensteinAA{
  D.~Berenstein,
  JHEP {\bf 0601}, 125 (2006)
  [arXiv:hep-th/0507203].
}

\lref\MandalTK{
  G.~Mandal and N.~V.~Suryanarayana,
  ``Counting 1/8-BPS dual-giants,''
  arXiv:hep-th/0606088.

}

\lref\BasuID{
  A.~Basu and G.~Mandal,
  arXiv:hep-th/0608093.
}

\lref\BasuID{
  A.~Basu and G.~Mandal,
  ``Dual giant gravitons in AdS(m) x Y**n (Sasaki-Einstein),''
  arXiv:hep-th/0608093.
}

\lref\MinwallaRP{
  S.~Minwalla,
  ``Particles on AdS(4/7) and primary operators on M(2/5) brane
  worldvolumes,''
  JHEP {\bf 9810}, 002 (1998)
  [arXiv:hep-th/9803053].
}

\lref\GunaydinTC{
  M.~Gunaydin and N.~P.~Warner,
  ``Unitary Supermultiplets Of Osp(8/4,R) And The Spectrum Of The S(7)
  Compactification Of Eleven-Dimensional Supergravity,''
  Nucl.\ Phys.\ B {\bf 272}, 99 (1986).
}

\lref\CasherYM{
  A.~Casher, F.~Englert, H.~Nicolai and M.~Rooman,
  ``The Mass Spectrum Of Supergravity On The Round Seven Sphere,''
  Nucl.\ Phys.\ B {\bf 243}, 173 (1984).
}

\lref\GunaydinWC{
  M.~Gunaydin, P.~van Nieuwenhuizen and N.~P.~Warner,
  ``General Construction Of The Unitary Representations Of Anti-De Sitter
  Superalgebras And The Spectrum Of The S**4 Compactification Of
  Eleven-Dimensional Supergravity,''
  Nucl.\ Phys.\ B {\bf 255}, 63 (1985).
}

\lref\AharonyRM{
  O.~Aharony, Y.~Oz and Z.~Yin,
  ``M-theory on AdS(p) x S(11-p) and superconformal field theories,''
  Phys.\ Lett.\ B {\bf 430}, 87 (1998)
  [arXiv:hep-th/9803051].
}

\lref\MinwallaKA{
  S.~Minwalla,
  ``Restrictions imposed by superconformal invariance on quantum field
  theories,''
  Adv.\ Theor.\ Math.\ Phys.\  {\bf 2}, 781 (1998)
  [arXiv:hep-th/9712074].
}

\lref\KinneyEJ{
  J.~Kinney, J.~M.~Maldacena, S.~Minwalla and S.~Raju,
  ``An index for 4 dimensional super conformal theories,''
  arXiv:hep-th/0510251.
}

\lref\AharonyAN{
  O.~Aharony, M.~Berkooz and N.~Seiberg,
  ``Light-cone description of (2,0) superconformal theories in six
  dimensions,''
  Adv.\ Theor.\ Math.\ Phys.\  {\bf 2}, 119 (1998)
  [arXiv:hep-th/9712117].
}

\lref\ggsus{
  J.~McGreevy, L.~Susskind and N.~Toumbas,
  ``Invasion of the giant gravitons from anti-de Sitter space,''
  JHEP {\bf 0006}, 008 (2000)
  [arXiv:hep-th/0003075].
}

\lref\ggrob{
  M.~T.~Grisaru, R.~C.~Myers and O.~Tafjord,
  ``SUSY and Goliath,''
  JHEP {\bf 0008}, 040 (2000)
  [arXiv:hep-th/0008015].
}

\lref\ggaki{
  A.~Hashimoto, S.~Hirano and N.~Itzhaki,
  ``Large branes in AdS and their field theory dual,''
  JHEP {\bf 0008}, 051 (2000)
  [arXiv:hep-th/0008016].
}

\lref\ggsumit{
  S.~R.~Das, A.~Jevicki and S.~D.~Mathur,
  ``Vibration modes of giant gravitons,''
  Phys.\ Rev.\  D {\bf 63}, 024013 (2001)
  [arXiv:hep-th/0009019].
}

\lref\ggouyang{
  P.~Ouyang,
  ``Holomorphic D7-branes and flavored N = 1 gauge theories,''
  Nucl.\ Phys.\  B {\bf 699}, 207 (2004)
  [arXiv:hep-th/0311084].
}

 \vskip 0.8cm

\Title{\vbox{\baselineskip12pt\hbox{hep-th/yymmddd}}}
{\vbox{\centerline{Supersymmetric States in M5/M2 CFTs}}}

 \centerline{Sayantani Bhattacharyya and  Shiraz
Minwalla}
\smallskip

\centerline{Department of Theoretical Physics,} \centerline{Tata
Institute of Fundamental Research, Homi Bhabha Rd, Mumbai 400 005.}

\vskip .6in \centerline{\bf Abstract}{We propose an exact, finite
$N$ formula for the partition function over $1/4^{th}$ BPS states in
the conformal field theory on the world volume of $N$ coincident
$M5$ branes, and $1/8^{th}$ BPS states in the theory of $N$
conincident $M2$ branes. We obtain our partition function by
performing the radial quantization of the Coulomb Branches of these
theories and rederive the same formula from the quantization of
supersymmetric giant and dual giant gravitons in $AdS_7 \times S^4$
and $AdS_4 \times S^7$. Our partition function is qualitatively
similar to the analogous quantity in ${\cal N}=4$ Yang Mills. It
reduces to the sum over supersymmetric multi gravitons at low
energies, but deviates from this supergravity formula at energies that 
scale like a positive power of $N$. }

\smallskip
\Date{} 


\newsec{Introduction}

The low energy dynamics on the world volume of $N$ coincident $M5$
or $M2$ branes is governed by maximally supersymmetric conformal
field theories in $d=6$ and $d=3$ dimensions. The AdS/CFT
correspondence `solves' these theories at large $N$. While rather
little is understood about these theories at finite $N$ ($N \neq
1$), it is known that they possess exactly flat directions along a
Coulomb branch; a reflection of the ability of these $N$ parallel
branes to separate along transverse directions. This Coulomb branch
is the metrically flat space $(R^8)^N/S_N$ or $(R^5)^N/S_N$, see
\SeibergAX; the quotient by the symmetric group $S_N$ reflects the
identical nature of these branes.

In this note we study the radial quantization of (a sub class of)
these Coulomb branch solutions; we pause to explain what this means.
Quantum field theories on $R^d$ are most often quantized by
associating Hilbert Spaces with field configurations along constant
time $R^{d-1}$ slices. Under this procedure, the distinct M2 and M5
branes Coulomb branch solutions parameterize distinct superselection
sectors. In this note we instead study the radial quantization of
the worldvolume theories of M5 and M2 branes on $R^6$ and $R^3$
respectively. This procedure is equivalent to the quantization of
these theories on $S^{5} \times R$ or $S^2 \times R$, and is natural
from several points of view. First, it introduces a mass gap into
the system, regulating potentially severe infrared divergences.
Relatedly (and more importantly for this note) it yields the dual to
$M$ theory on global, geodesically complete, $AdS_4 \times S^7$ and
$AdS_7 \times S^4$ respectively.

Under radial quantization the world volume theories we study each
have a unique vacuum. Distinct Coulomb branch configurations are
normalizable, finite energy fluctuations about this vacuum
\foot{They map to time dependent solutions of the relevant theories
on $S^{d-1} \times R$.}. Sub classes of these solutions are
respectively ${1 \over 8}$ and ${1 \over 4}$ BPS. In this note we
(radially) quantize these supersymmetric solutions and compute the
partition functions $Z^N_{5}$ and $Z_2^N$, over the resultant
Hilbert Space. We conjecture \foot{See \AharonyAN\ and \BasuID\ for
related earlier work.} that $Z^N_{5}$ and $Z_2^N$ are the exact,
finite $N$ partition function over the 1/8 or 1/4 BPS Hilbert space
of the theories we study \foot{Equivalently, we conjecture that the
Coulomb Branch exhausts the set of appropriately supersymmetric
`classical' configurations in the full non abelian theory. In
Section 3 we demonstrate that the analogeous result is true for the
low energy theory of D3-branes in IIB theory, i.e. ${\cal N}=4$ Yang
Mills theory.}. Our results agree in particular with the spectrum of
`single trace' chiral primaries for the (0,2) theory of the M5 brane
computed in \AharonyAN.

We proceed to employ the AdS/CFT correspondence to gather evidence
for our conjecture. In particular we demonstrate that

\item{1.} In the large $N$ limit $Z^N_5$ and $Z^N_2$ reduce
to the partition function over supersymmetric multi graviton
configurations in $AdS_7 \times S^4$ and $AdS_4 \times S^7$
respectively.

\item{2.} The partition function obtained from the quantization
of the full manifold of Mikhailov's supersymmetric giant gravitons
(\refs{\andrei, \ggsus} in $AdS_7 \times S^4$ and $AdS_4 \times S^7$
(along the lines of earlier studies \refs{\BeasleyXV, \our} agrees
exactly with $Z^N_{5}$ and $Z^N_2$. We also argue that the
quantization multi dual giant gravitons may be regarded as bulk
analogue of the quantization of the Coulomb Branch. A direct study
of the quantization of dual giant gravitons (following previous
studies \refs{\MandalTK,\BasuID}) yields additional support for our
formula for $Z^N_{5}$ and $Z^N_2$.

The rest of this note is organized as follows. In section 2 we
briefly review the supersymmetry algebra of the M5 and  M2 brane
world volume theories. We give a precise characterization of the
supersymmetric states and the partition functions $Z_{5}$ and $Z_{2}^N$ that
we study in this note. We also compute the partition function over
supersymmetric multi gravitons in $AdS_7 \times S^4$ and $AdS_4
\times S^7$. In section 3  we compute the partition function
$Z_{5}^N$ and $Z^N_2$ by quantizing the relevant Coulomb branches.
We argue that our procedure for obtaining $Z_{5}^N$ and $Z_2^N$ has
a close bulk analogue in the quantization of $N$ non interacting
dual giant gravitons along the lines of \refs{\MandalTK,\BasuID}. In
Section 4 we compute the partition function over appropriately
supersymmetric giant gravitons in $AdS_7 \times S^4$ and $AdS_4
\times S^7$, using the methods of \refs{\BeasleyXV, \our}. Though
this procedure is atleast superficially quite different from our
quantization of the Coulomb branch of section 3, it gives exactly
the same result. In section 5 we end with a brief discussion.

\newsec{Supersymmetric States and the Supergravity Partition Function}

In this section we group theoretically characterize the
supersymmetric states of interest to us in this note\foot{See for
instance \MinwallaKA\ for information about these superalgebras and
their unitary representations.}.  We also use the AdS/CFT
correspondence to compute the partition function over these states
in the large $N$ limit.

\subsec{M5 branes}

The bosonic part of the supersymmetry algebra of the $(0,2)$ theory
on the world volume of the M5 brane is $SO(6,2) \times SO(5)$.
Supersymmetry generators ($Qs$) are simultaneously spinors of
$SO(5)$ and chiral spinors of $SO(6) \in SO(6,2)$. The Hermitian
conjugates of these operators ($Ss$) are also $SO(5)$ spinors and
antichiral $SO(6)$ spinors. Consider the set of four $Qs$ with given
a given set of $SO(5)$ charges. For concretenes we choose these
$SO(5)$ charges to be $H_1=H_2= \half$ where $H_1$ and $H_2$ are
Cartan generators of $SO(5)$ that generate rotations in orthogonal
planes in an embedding $R^5$ and arbitrary $SO(6)$ charges. These
four supersymmetries, together with their Hermitian conjugates,
generate the compact superalgebra $SU(4/1)$.

We are interested in the $SU(4/1)$ invariant states in the $(0,2)$
theory of $N$ coincident M5-branes. There exists an equivalent
characterization of these states; they are $SO(6)$ singlets that
simultaneously obey the BPS bound \eqn\bpsf{E=2H_1+2 H_2.} In this
note we will propose a formula for \eqn\defpf{Z_{5}^N=\Tr e^{-\mu_1
H_1-\mu_2 H_2}} where the trace is taken over all $SU(4/1)$
invariant states in the theory of $N$ coincident M5 branes.

Let us immediately compute this partition function in the limit of
infinite $N$. Recall that the Maldacena dual of the $(0,2)$ theory
is M theory on $AdS_7 \times S^4$. In the infinite $N$ limit all
finite energy states are non interacting multi supergravitons. The
spectrum of gravitons in $AdS_7\times S^4$ was worked out and
arranged into multiplets of the superconformal group in \GunaydinWC\
and references therein (see also \refs{\MinwallaRP, \AharonyRM}).
All gravitons appear in short representations of the superconformal
algebra. The primaries of these representations are $SU(4)$
singlets, tranform in the $n^{th}$ symmetric traceless
representation of $SO(5)$, and have energy equal to $2n$. It turns
out that the only $SU(4/1)$ invariant states in these
representatations are those primary states with charges that obey
the equation $H_1+H_2=n$. There exists one such state for every
partitioning of $n$ into a sum of two non negative integers. Summing
over all primaries (and hence over all $n$) we conclude that the set
of supergravitons of interest to us are labeled by two non negative
integers $(n_1, n_2)$ that cannot simultaneously be zero; the
corresponding states have charges $(H_1, H_2)=(n_1, n_2)$. The large
$N$ fixed chemical potential limit of the $SU(4/1)$ invariant
subsector of the $(0,2)$ theory is simply the Fock space over these
supersymmetric graviton states. The partition function over this
Fock space is given by \eqn\gravf{Z^\infty_5= \Tr_{{\cal H}_\infty}
e^{-\mu_1 H_1-\mu_2 H_2} =\prod_{n_1, n_2=0}^\infty {1 \over 1-
e^{-n_1 \mu_1 -n_2 \mu_2}}} where the case $n_1=n_2=0$ is excluded
from the product.

\subsec{M2 Branes}

The bosonic part of the supersymmetry algebra of the world volume of
the M2 brane is $SO(3,2) \times SO(8)$. Supersymmetry generators
($Qs$) and their complex conjugates $S s$ are both simultaneously
spinors of $SO(3)\in SO(3,2)$ and chiral spinors of $SO(8)$.
Consider the set of 2 $Qs$ with arbitrary $SO(3)$ charges and a
given set of $SO(8)$ charges. For concreteness we choose these to be
$H_1=H_2=H_3=H_4= \half$ where $H_i$, $i= 1 \ldots 4$ are Cartan
generators of $SO(8)$ that generate rotations in mutually orthogonal
planes in $R^8$. These two supersymmetries, together with their
Hermitian conjugates, generate the compact superalgebra $SU(2/1)$.
$SU(2/1)$ invariant states admit an alternate characterization; they
are $SO(3)$ singlets that simultaneously obey the BPS bound
\eqn\bpst{E=\half \left( H_1+ H_2 + H_3+H_4 \right).} In this note
we propose a formula for \eqn\defpt{Z_{2}^N=\Tr e^{-\sum_{i=1}^4
\mu_i H_i}} over all such states.

The Maldacena dual of the M2-brane theory is M theory on $AdS_4
\times S^7$.  The spectrum of gravitons on $AdS_4 \times S^7$ was
worked out in  \CasherYM. This spectrum was arranged in
representations of the superconformal algebra in \GunaydinTC\ (see
also \refs{\MinwallaRP, \AharonyRM}). All gravitons lie in
supershort representations of the superconformal algebra. The
primaries of these representations are $SO(3)$ scalars, transform in
traceless symmetric tensors (of arbitrary rank $n$ ) of $SO(8)$ and
have energy given by ${n \over 2}$. It turns out that the only
$SU(2/1)$ invariant states in these multiplets are those primary
states whose charges obey $\sum_{i=1}^4 H_i=n$. Suming over all
primaries (and so all $n$) we conclude that the set of $SU(2/1)$
invariant gravitons on this space is labeled by four non negative
integers that cannot all be zero. In terms of these integers
$(H_1,H_2, H_3, H_4)= (n_1, n_2, n_3, n_4)$ where $n_i$ ($i =1
\ldots 4$). It follows that the partition function over multi super
gravitons is given by \eqn\gravt{Z^\infty_2= \Tr_{{\cal H}_\infty}
e^{-\mu_1 H_1-\mu_2 H_2 -\mu_3 H_3 -\mu_4 H_4} =\prod_{n_1, n_2,
n_3, n_4=0}^\infty {1 \over 1- e^{-n_1 \mu_1 -n_2 \mu_2 -n_3 \mu_3
-n_4 \mu_4}}} where the term with all $n_i$ zero is excluded from
the product.

\newsec{Quantization of the Coulomb Branch}

In this section we determine the partition function $Z^N_{5}$ and
$Z_2^N$ by radially quantizing Coulomb branch configurations of the
world volume theories of $M$5 and $M$2 branes. We begin this section
with a brief discussion of the radial quantization of conformally
coupled scalar field theories in arbitrary dimension.

\subsec{Radial Quantization of a Free Scalar Fields in Aribtrary Dimension}

Consider a free, complex, conformally coupled $d$ dimensional
 scalar field on a unit $S^{d-1}$
\eqn\act{S=\int dt d^{d-1}x  \left( \del \phi_S . \del \phi_S^* -
{d-2 \over 2 } ~\phi_S \phi_S^* \right).} where $\phi_S^*$ is the
complex conjugate of $\phi_S$. Analytically continuing to Euclidean
time and conformally mapping to the $R^{d}$ we find the Euclidean
action \eqn\fcs{S=\int
\partial {\bar \phi} . \partial \phi } where
\eqn\whatph{\phi(x)={\phi_S(x) \over |x|^{{d-2 \over 2}}},
~~~~~{\bar \phi}(x^\mu)= {\phi^*_S({x^\mu \over x^2}) \over
|x|^{{d-2 \over 2}}}.} Notice that \eqn\fieldrf{{\bar \phi}(x)=
|x|^{-({d-2})} \phi^*(x^\mu/|x|^2); } in particular  ${\bar \phi}$
is not simply the complex conjugate of $\phi$. It is possible to
check that the equations of motion  \eqn\eom{
\partial^2 \phi=
\partial^2 {\bar \phi}= 0.}
and \fieldrf\ are mutually consistent. \foot{This follows from the
observation that $\partial_x^2 (y^{d-2} \chi)= y^{d+2} \partial_y^2
\chi$ where $\partial_x^2$ and $\partial_y^2$ are the  Laplacian
with respect to the $x$ and $y^\mu=x^\mu/|x|^2$ respectively.}

Regular solutions of \act\ map to field configurations $\phi$ and
${\bar \phi}$ whose singularities are localized at $x=0$ and
$x=\infty$. \foot{The coefficients of solutions regular at zero turn
into creation operators , while the coefficients of solutions
regular at infinity turn into destruction operators upon
quantization. } The general solution of \eom\ and \fieldrf\ is given
by \eqn\solneom{ \eqalign{ \phi&= \sum c^*_{\mu_1 ... \mu_m}
x^{\mu_1} ... x^{\mu_m} + y^{d-2} \sum d_{\nu_1 ... \nu_n} y^{\nu_1}
... y^{\nu_n} \cr {\phi}&= \sum d^*_{\nu_1 ... \nu_m} x^{\nu_1} ...
x^{\nu_n} + y^{d-2}  \sum c_{\mu_1 ... \mu_m} y^{\mu_1} ...
y^{\mu_n} \cr {\rm where}~~~y^\mu &={x^\mu \over x^2} }} and
$c_{\mu_1...\mu_m}$ and $d_{\mu_1 ... \mu_m}$ are arbitrary complex
traceless symmetric tensors.

It follows from \fieldrf\ (and we note for future reference) that
\eqn\dcc{\eqalign{ \partial_\mu{\bar \phi}&=
({1\over {x^{d-2}}})[-(d-2) {{x_\mu}\over x^2} \phi^* - 2 {{x_\mu x^\alpha
\partial_\alpha \phi^*}\over x^4} +{{\partial_\mu \phi^*}\over x^2}] \cr
x.\partial {\bar \phi} &= -({1\over {x^{d-2}}})[(d-2) \phi^*
+{{x. \partial \phi^*}\over x^2}] \cr
\partial \phi . \partial {\bar \phi}&= ({1\over {x^{d-2}}})
[-(d-2){{ x. \partial \phi \phi^*}\over x^2}
- 2 {{(x. \partial \phi) (x. \partial \phi^*)}\over x^4}
+ {{\partial \phi. \partial \phi^* }\over x^2}].}}

The symplectic form from \act\  \eqn\sa {\omega  = \int_{S^{(d-1)}}
\left[d\dot{\phi_S} \wedge d{\phi_S^*} + d{\dot{{\phi_S^*}}} \wedge
d{\phi_s} \right]} translates into \eqn\sb {\eqalign{-i\omega=&
\int_{S^{d-1}} d \left[(x.\partial) \left(x^{{(d-2)}\over 2}
\phi(x^\mu) \right) \right]\ \wedge\ d\left[x^{{d-2}\over
2}\phi^*({{x^{\mu}}\over x^2}) \right]\cr & +
d\left[(x.\partial)\left(x^{-{(d-2)}\over 2}\phi^*({{x^{\mu}}\over
x^2}) \right)\right] \  \wedge \ d\left[x^{{(d-2)}\over
2}\phi(x^\mu)\right].}}

The generator of time translations in \act\ maps to the generator of
scale transformations in \fcs. The corresponding conserved current
is given by
\eqn\scaletransf{D_\mu = x_\mu
\left( {\partial {\bar \phi} . \partial \phi } \right)
- {{(d-2)}\over 2}[\partial_\mu \phi  {\bar \phi} + \partial_\mu {\bar \phi} \phi ]
- \partial_\mu \phi (x. \partial {\bar \phi}) -
\partial_\mu \phi (x. \partial {\bar \phi}), }
and its conserved charge $E$  (the energy) is given by
\eqn\formd{\eqalign{ E&= \int |x|^{(d-2)} x.D d\Omega_{(d-1)} \cr &=
\int
\partial \phi. \partial {\bar \phi} -2  (x. \partial \phi) (x.
\partial {\bar \phi}) - {d-2 \over 2} [x . \partial \phi {\bar \phi}
+ x.{\partial {\bar \phi}} \phi] \cr &=\int \partial \phi . \partial
\phi^*  + ({{d-2}\over 2})[(x. \partial \phi) \phi^* + (x. \partial
\phi^*) \phi] + {{(d-2)^2}\over 2} \phi \phi^*  }} where the
integrals in the last two lines are evaluated on the unit $S^{d-1}$
$|x|=1$.

The action \fcs\ is invariant under a $U(1)$ scaling of the field $\phi$.
The corresponding Noether current, $H_\mu$ is given by
\eqn\chargecur{H_\mu= {\bar \phi} \partial_\mu \phi - \phi
\partial_\mu {\bar \phi}.}
The associated conserved charge $H$ is given by \eqn\uocs{\eqalign{
H&= \int d\Omega_{(d-1)} x^{(d-2)} x.H \cr &= \int  \phi^* x.
\partial \phi + \phi x. \partial \phi^* + (d-2) \phi \phi^*} } where
the integral in the last line is evaluated on the on the unit sphere
$|x|=1$.

Note that \eqn\bps{\eqalign{ E-{{(d-2)}\over 2}H &= \int \partial
\phi . \partial \phi^* \cr
E+{{d-2}\over 2}H&= \int \partial \phi .
\partial \phi^*  + (d-2)[(x. \partial \phi) \phi^* + (x. \partial
\phi^*) \phi] + (d-2)^2 \phi \phi^*\cr 
&=  \int \partial \bar{\phi}. \partial \bar{\phi}^* } }
It follows that $(E\pm {{(d-2)}\over 2}H)$ is
zero if and only if respectively $\phi$ or ${\bar \phi}$ are constant. 
Further 
\eqn\moreen{E=\half \int \left( \partial
\phi . \partial \phi^* + \partial \bar{\phi}. \partial \bar{\phi}^* \right).
} Consequently, in order for a configuration to have zero energy, $\phi$ 
and ${\bar \phi}$ must each be constant. However it follows from \fieldrf\ 
that $\phi$ and ${\bar \phi}$ can both be constant only if they are both 
zero; thus  $\phi=\bar \phi=0$ is the only zero energy 
configuration.

Using \sa, \scaletransf\ and \chargecur\ it is a simple matter to
quantize the general solution \solneom\ to obtain the spectrum of
energies and charges. We illustrate this on a subclass of solutions.
Let $z$ represent any complex direction in $R^d$ and consider the
subclass of solutions \eqn\csfield{\phi= \sum_n {c^*_n \over
\sqrt{K_n}} z^n.} where the constant $K_n$ is chosen (for
convenience) to be \eqn\kndef{ \eqalign{K_n=& 4 \pi n \Omega_{d-3}
\int_0^1 r^{2n-1}(1-r^2)^{{d-4}\over 2} dr \cr  =& 2 \pi (d-2+2n)
\Omega_{d-3} \int_0^1 r^{2n+1}(1-r^2)^{{d-4}\over 2} dr.}} We find
\eqn\sba{ \eqalign{ H &= \sum_n |c_n|^2 \cr E-{d-2 \over 2}H & =
\sum_n n |c_n|^2 \cr \omega & =  i\sum_n dc^*_n \wedge dc_n }} The
quantization of \sba\ turns $c_n^*$ and $c_n$ respectively into
creation and annihilation operators of unit charge and energy
$n+{d-2 \over 2}$ respectively.

The quantization of the general solution \solneom\ proceeds along
similar lines.  $c_{\mu_1 ... \mu_n}$ and $c^*_{\mu_1 ... \mu_n}$
turn into operators that are proportional to unit normalized
creation and annihilation operators of energy $n+{d-2 \over 2}$ and
unit $U(1)$ charge under quantization.  Similarly $d_{\mu_1 ...
\mu_m}$ and $d^*_{\mu_1 ... \mu_m}$ become creation and annihilation
operators of energy $m+{d-2 \over 2}$ and negative unit $U(1)$
charge respectively.

Real scalar fields obey the constraint  \eqn\reality{\psi = \bar
\psi.} The quantization of these theories proceeds along similar
lines. The equation \moreen\ also applies to real scalar fields with 
an extra factor of $\half$ on the RHS in standard real normalization  (note
also  that the two terms on the RHS of this equation are equal in this 
special case). As it is impossible for a nonzero constant value of 
$\psi$ to obey  \reality\ (see \fieldrf) it follows that $\psi=0$ is the 
only field configuration with vanishing energy. 

\subsec{Quantization of the BPS sector of the U(1) theory}

A single M2 brane has four free complex scalar fields, $\phi_i$ $(i
= 1 \ldots 4)$ on its world volume. The field $\phi_i$ has charge
$\delta_i^j$ under the Cartan $U(1)$ rotations $H_j$ of section 2.
The BPS combinaton \bps\ evaluates to \eqn\bbpp{E-\half
{\sum_{i=1}^4} H_i =\sum_{i=1}^4 \int \partial \phi_i .
\partial \phi^*_i}
It follows that the constant field configurations $\phi_i(x)={\phi_i
\over \sqrt{2 \pi }}$ are the only classical BPS configurations; the
corresponding symplectic form, energy and charge formulae are
\eqn\sdnt{\eqalign{\omega &= i \sum_{i=1}^4 d\phi_i \wedge
d\phi_i^{*} \cr E &= {{1}\over 2} \sum_{i=1}^4 \phi_i \phi^{*}_i \cr
H_i &= {\phi_i \phi_i^{*} } .}} The quantization of \sdnt\ yields
the Hilbert space of a four dimensional harmonic oscillator; charge
operator $H_i$ turns into the number operator of the $i^{th}$
oscillator on this space.

A single M5 brane has two complex free scalar fields, $\phi_i$ $(i =
1 \ldots 2)$ and one real scalar field $\psi$ on its world volume.
The complex fields $\phi_i$ have charges $\delta_i^j$ under the
charge $H_j$; $\psi$ is uncharged. We have 
\eqn\bbpp{E-\half
{\sum_{i=1}^2} H_i = \half \partial \psi .
\partial \psi^* + \sum_{i=1}^2 \int \partial \phi_i .
\partial \phi^*_i.}
The set of classically BPS
configurations is given by constant $\phi^i(x)={\phi^i
\over \sqrt{(2 \pi)^3}}$ and $\psi=0$ (recall that it is impossible  
for a nonzero constant $\psi$ to obey \reality). On this class of solutions
\eqn\sdnf{\eqalign{\omega &= i \sum_{i=1}^2 d\phi_i \wedge
d\phi_i{*} \cr E &= 2 \sum_{i=1}^2 \phi_i \phi_i{*} \cr H_i &
={\phi_i \phi_i^{*}} }} The quantization of \sdnf\ yields the
Hilbert space of a two dimensional harmonic oscillator; the charge
operator $H_i$ is the number operator of the $i^{th}$ oscillator on
this space.

In summary, the symplectic manifold of classically SU(4/1) and
SU(2/1) invariant configurations on the theory of a single M5 or M2
brane is $C^2$ and $C^4$ respectively. The quantization of these
manifolds yields the the Hilbert space of two and four dimensional
harmonic oscillators, respectively.

\subsec{Quantization of the Coulomb Branch}

As we have remarked in the introduction, the world volume theory of
$N$ M5 or M2 branes possesses Coulomb branches. Away from their
singularities, these Coulomb branches are metrically flat; the
spaces in question are  $(R^8)^N/S_N$ and $(R^5)^N/S_N$
respectively. Using the results of the previous section, it follows
that classically $SU(4/1)$ or $SU(2/1)$ invariant Coulomb branch
solutions constitute the symplectic manifolds $(C^4)^N/S_N$ and
$(C^2)^N/S_N$ respectively. The quantization of these manifolds
yields fock space of $N$ identical, non interacting bosons in a 4 or
2 dimensional harmonic oscillator potential. The partition function
for this system follows immediately from the usual formulas of Bose
Statistics;  \eqn\pfmf{ \sum_{m=1}^\infty p^N Z_5^N = \prod_{n_1,
n_2=0}^\infty {1 \over 1-p e^{-n_1 \mu_1 -n_2 \mu_2}}} and
\eqn\pfmt{ \sum_{m=1}^\infty p^N Z^N_5 = \prod_{n_1, n_2, n_3,
n_4=0}^\infty {1 \over 1-p e^{-n_1 \mu_1 -n_2 \mu_2 -n_3 \mu_3 -n_4
\mu_4}}}

\subsec{Absence of additional Supersymmetric Configurations in
${\cal N}=4$ Yang Mills}

As we have explained in the previous subsection, the world volume
theory of M5 and M2 branes has a class of, respectively,  SU(4/1)
and SU(2/1) invariant configurations on their Coulomb branch. We
conjecture that these configurations are exhaustive; that the full
non ablelian world volume theories have possess no further SU(4/1)
or SU(2/1) invariant configurations, so that \pfmt\ and \pfmf\
represent the exact partition functions over SU(4/1) and SU(2/1)
invariant states in these theories.

Our poor understanding of the structure of the non abelian theory on
M2 and M5 branes makes it difficult to directly verify this
conjecture. In this sub section we will, however, demonstrate that
the analogous claim is indeed true of much better understood  world
volume theory of $N$ D3 branes - $U(N)$ ${\cal N}=4$ Yang Mills
theory.

The $1/8^{th}$ BPS sector of ${\cal N}=4$ Yang Mills theory has
recently been studied in some detail in \refs{\KinneyEJ, \our,
\MandalTK}; we will not pause here to characterize this subsector
group theoretically, but instead refer to reader to \KinneyEJ\ for
such details. This worldvolume theory possesses 3 complex adjoint
valued scalar fields $\phi_i= \psi_{2i-1}+ i \psi_i$ (where $\psi_i,
~~~i=1 \ldots 6$ are Hermitian $N\times N$ matrix valued scalar
fields). The restriction of this sector to states (or operators)
made entirely out of scalars\foot{It should be possible to construct the 
full ${1 \over 8}^{th}$ BPS cohomology of Yang Mills theory 
(studied in \KinneyEJ) from the radial quantization of supersymmetric 
fermionic field configurations, together with the scalars studied in this 
paper. We expect the relevant fermionic configurations to be constant 
diagonal gaugino fields. As these configurations have no analogues in the 
theory of M5 and M2 branes, we do not perform a detailed study of these 
configurations in this paper. } analogous, in many ways, to the
supersymmetric states we have been studying on M5 and M2 brane world
volumes; this subclass of $1/8^{th}$ states consist of Lorentz
scalars that obey the BPS bound $E={4-2 \over 2}(H_1+H_2+H_3) =
H_1+H_2+H_3$ where $H_i$ refer to the three Cartan generators of the
$SO(6)$ R symmetry of this theory. The Euclidean Lagrangian of this
theory - restricted to configurations over which only scalar fields
are nonzero - is given by \eqn\lag{{1 \over g_{YM}^2} \int \Tr
\left(\sum_i \partial \phi_i
\partial {\bar \phi}_i + \half \sum_{i \neq j} [\psi_i,
\psi_j][\psi_j, \psi_i] \right) } (recall $\psi_i$ are the real
components of the complex scalars $\phi_i$). Performing the radial
quantization of this Lagrangian, imitating the work out of
subsection 3.1, we find \eqn\bpsbn{E-\sum_{i=1}^3 H_i={1 \over
g^2_{YM}} \Tr \int \left( \sum_{i=1}^3
\partial \phi_i . \partial \phi_i^\dagger  + \sum_{m, n=1}^6
[\psi_m, \psi_n][\psi_n, \psi_m] \right) .} It follows  from \bpsbn\
that the set of scalar BPS configurations of this theory consist of
constant diagonal matrices $\phi_i$, further gauge invariance
requires us to identify matrices with permuted eigenvalues
\foot{Mapping to $S^3$ we recover a slight generalization of the
diagonal solutions considered in \ggaki.}  (see \BerensteinAA\ for 
closely related remarks). This set of matrices
parameterizes the Coulomb branch studied above. We conclude that the
set of 1/8 BPS configurations in ${\cal N}=4$ Yang Mills lies
entirely within the Coulomb branch \foot{The quantization of these
configurations may be carried out; the relevant formulae are
\eqn\sdnt{\eqalign{\omega &= i \sum_{i=1}^3 \Tr d\phi_i \wedge
d\phi_{i}^{\dagger} \cr E &= \sum_{i=1}^3 \Tr \phi_i \phi_{i}^
\dagger \cr H_i &= \Tr \phi_i \phi_i^{\dagger} }} }; in other words
the diagonal supersymmetric configurations that generalize those of
\ggaki\ constitutes the full set of scalar $1/8^{th}$ BPS
configurations of Yang Mills theory.

\subsec{Dual Giants: Bulk duals of the Coulomb Branch}

The bulk $AdS$ dual of an M5 or M2 brane on its Coulomb Branch is an
M5 or M2 brane that is puffed out in $AdS_7$ or $AdS_4$ in an  SO(6)
or SO(3) invariant manner. The SO(6) or SO(3) invariant
configurations in $AdS_7$ or $AdS_4$ are a one parameter set of $5$
and $2$ dimensional spheres that foliate spatial sections of $AdS$.
\foot{Exceptional orbits consist of  points (fixed points under the
rotational action). These orbits may be regarded as the zero radius
limit of our spheres, and are automatically taken into account
below.} Branes wrapping such spheres,  and otherwise moving on $S^4$
or $S^7$ in a supersymmetric fashion, have been identified
\refs{\ggaki, \ggrob} and studied in detail; they are called dual
giant gravitons. In this subsection we review the construction of
these  configurations \refs{\ggrob, \ggaki} and their quantization
(partially performed in \BasuID; see also \refs{\ggsumit, \ggouyang,
\MandalTK} for related work). We find that the result of the
quantization of a single supersymmetric dual giant graviton is
identical to the quantization of a single brane on the Coulomb
Branch discussed in the previous section. Our discussion of the
quantization of dual giant gravitons overlaps with the  in the case
of the M2 brane..

Let us first describe dual giant gravitons in $AdS_{m+2} \times
S^{n+2}$ \refs{\ggrob, \ggaki}. Although backgrounds of this form
are known to appear in string theory only for specific values of $m$
and $n$, in this section and the next we will leave $m$ and $n$
arbitrary. The formulae we present below apply to M5, M2, and D3
brane dual giant gravitons upon setting $(m,n)$ to $(5,2)$, $(2,5)$
and $(3,3)$ respectively.

Let the metric on $AdS_{m+2} \times S^{n+2}$ be given by
\eqn\metonspace{d{\tilde s}^2= R_1^2 \left( -\cosh^2 \rho dt^2 + d
\rho^2 + \sinh^2 \rho d \Omega_{m}^2 \right) + R_2^2 d
\Omega_{n+2}^2} where $\Omega_k$ is the metric on the unit $k$
sphere. Below we will find it convenient to work with the scaled
metric \eqn\scaledmet{ ds^2={d {\tilde s}^2 \over R_1^2} =G_{\mu
\nu}dx^\mu dx^\nu  .} $R={R_2 \over R_1}$ is the radius of the
sphere in the new rescaled metrics \scaledmet; in the solutions that
appear in string theory $R={2 \over m-1}={2 \over d-2}$ where $d$ is
the spacetime dimension of the dual world volume theory. In
particular $R=\half$ in $d=6$ and $R=2$ in $d=3$.  The space
\metonspace\ has a nonzero $n+1$ form gauge potential $A_{n+1}$
turned on such that the corresponding field strenth preserves the
symmetry of $S^{n+2}$ and $\int_{S^{n+2}} d A_{n+1}= 2 \pi N$. We
denote the $m+1$ form dual to $A_{n+1}$ by $A_{m+1}$.

The action for an $m$ brane propagating in this background is given
by \eqn\actmb{S=-{N^{m-1 \over 2} 2^{{m-3 \over 2}} \over \Omega_m}
\int \sqrt{-g} + \int A_{m+1}.} For furture use we note that the
action for an n brane propagating in the same background is given by
\eqn\actnb{S=-{N\over R^{n+1} \Omega_n}\int \sqrt{-g} + \int
A_{n+1}.} $g_{\alpha \beta}$ is the pullback of the metric $G_{\mu
\nu}$ (see \scaledmet) on the world volume of the brane in question.

As we have explained above, in this subsection we are interested in
an m-brane that completely wraps the $m$ dimensional sphere in
$AdS_{m+2}$ and so is effectively a particle in the remaining $n+4$
spacetime dimensions. The degrees of freedom of this particle are
the $n+2$ position coordinates in $S^{n+2}$ and its location in the
radial $\rho$ direction in $AdS_{m+2}$.

We will find it convenient to make a particular choice of
coordinates on $S^{n+2}$. Regarding $S^{n+2}$ as the unit sphere in
$R^{n+3}$ we set \eqn\cordsonsphere{x_1+ix_2= r_1 e^{i \theta_1},
~~~ \ldots,~ x_{n+2}+ix_{n+3}= r_{{n+3\over 2}} e^{i \theta_{{n+3
\over 2}} } } when $n$ is odd and \eqn\cordsonsphere{x_1+ix_2= r_1
e^{i \theta_1}, ~~~ \ldots x_{n+1}+ix_{n+2}= r_{{n+2 \over 2}} e^{i
\theta_{{n+2 \over 2}}}, ~~~ x_{n+3} = h } when $n$ is even. The
metric on $S^{n+2}$ is\eqn\metonsph{ \eqalign{ d\Omega_{n+2}^2 =&
dh^2 + \sum _{i=1}^{k}{dr_i}^2 + \left( d\sqrt{1-h^2-\sum_{i=1}^k
r_i^2} \right)^2 \cr &+ \left( 1-\sum_{i=1}^{k}{r_i}^2 - h^2 \right)
{d\theta_{k+1}}^2 + \sum_{i=1}^{k}{r_i}^2 {d\theta _i}^2 }} where $k
={n+1 \over 2}$ when $n$ is odd (in which case we simply set $h$ to
zero in \metonsph) and $k = {n \over 2}$ when $n$ is even.

The action for the dual giant graviton is given by
\eqn\bb{S=2^{{m-3}\over 2}N^{{m-1}\over 2} \int d\tau \  \cal L}
where \eqn\cb{ {\cal L}=+\dot t\sinh^{m+1}\rho -
        \sinh^m\rho \sqrt{{\dot t}^2\cosh^2\rho - \dot\rho^2
- R^2\left({{d\Omega_{n+2}}\over {d\tau}}\right)^2}} where
$({{d\Omega_{n+2}}\over {d\tau}})^2$ is the sigma model kinetic term
on a unit $S^{n+2}$ and, once again, $h$ is simply set to zero when
$n$ is odd.

In order to proceed we choose $\tau = t$. As the variables
$\theta_i$ do not appear in the lagrangian, $\dot {\theta_i}$ are
constants on all solutions to the equations of motion. Let us study
the ansatz $h$,  $r_i$,  ${{\dot \theta}_i}$ and $\rho=$ all
constant. This ansatz  yields solutions when $h=0$ (forced by the
$h$ equation of motion), ${\dot \theta}_i$ are all equal (forced by
the $r_i$ equation of motion) and ${\dot \theta}_i={1\over R}$
(forced by the $\rho$ equation of motion). Note that $r_i$, $\rho$
and the intial values of $\theta_i$ are unconstrained, and
parameterize distinct solutions. On solutions $H_i$ is the momentum
conjugate to $\theta_i$ while the energy is the negative of the
momentum conjugate to $t$; we find \eqn\hb{\eqalign{
\sum_{i=1}^{k+1}  H_i &= R (\sinh^{m-1}\rho) \sum_{i=1}^{k+1}
{r_i}^2 = R \sinh^{m-1} \rho \cr E & =  \sinh^{m-1} \rho = {1\over
R}\sum_{i=1}^{k+1} {H_i}.}} As a consequence of the last equation in
\hb\ the solutions described in the previous paragraph are all
`supersymmetric'.

     Notice that $h$ (which exist only for even $n$) is always zero on
solutions to the equations of motion. This is a dual expression of
fact that the real field $\psi$ of section 3.2 is zero on
`supersymmetric' Coulomb branch configurations.

On these solutions the momentum corresponding to $\r$ ,$r_i$ and $h$
all vanish. The momentum conjugate to $\theta_i$ is given by \eqn\db
{P_{\theta_i} = (\sinh^{m-1}\rho)R^2 {r_i}^2 \dot{\theta_i} = R
{r_i}^2  \sinh^{m-1}\rho.} It follows that \eqn\sympformdg{\omega=
\sum_i dP_{\theta_i} \wedge d \theta_i =  \sum_{i=1}^{k+1} 2R_i dR_i
\wedge d\theta_i = i\sum_{i=1}^{k+1} d \phi_i \wedge d \phi_i^*}
where \eqn\raddef{R_i = \sqrt{R} (\sinh^{{m-1}\over 2}\rho) r_i} and
\eqn\phidef{\phi_i=R_i e^{- \theta_i}.} The expressions for charges
are \foot{Similar results are reported  in \BasuID.}
\eqn\charen{\eqalign{ E&= {d-2 \over 2} \sum_i |\phi_i|^2 \cr
H_i&=|\phi_i|^2 .}} Notice that \sympformdg\ and \charen\ are
identical to \sdnf. It follows that the quantization of a single
dual giant graviton is identical to the quatization of a single
brane on the Coulomb branch. \foot{See \MandalTK\ for arguments that
suggest this identity should persist for collections of multi giant
gravitons.} The interesting translation formulas \raddef\ and
\phidef\ provide a dictionary to convert between spacetime
coordinates and Yang Mills field expectation values.

\newsec{Supersymmemtric States from Giant Gravitons}

In the previous section we have determined the partition function
over supersymmetric states on the M5 brane  or M2-brane world volume
by quantizing supersymmetric Coulomb branch configurations. In this
section we will demonstrate that there exists a bulk quantization -
superficially unrelated to the quantization of the Coulomb branch -
that permits relatively simple finite $N$ computation the spectrum
of supersymmetric states in the world volume theories of M5 and M2
branes. The procedure we refer to is the quantization of giant
gravitons.

Over 5 years ago Mikhailov constructed all $SU(4/1)$ invariant giant
graviton configurations in $AdS_7 \times S^4$ and $AdS_4 \times
S^7$. Mikhailov's giant gravitons all sit at the point $\rho=0$ in
$AdS_7$ or $AdS_4$ (i.e. at the fixed point of the $SO(6)$ or
$SO(3)$ killing symmetry in this space). The shape of these probes
with the $S^4$ or $S^7$ is described by an indirect construction.
Let us first consider the case of $S^4$. Let $S^4$ be embedded as
the unit sphere in $R^5$. The rotations $H_1$ and $H_2$ are
represented by killing vector fields corresponding to rotations in
orthogonal two planes in $R^5$. A rotation by angle ${\pi \over 2}$
simultaneously in each of the $H_1$ and $H_2$ directions yields a
complex structure in $C^2 \in R^5$. Let $z_1$ and $z_2$ be complex
coordinates in this $C^2$. Mikhailov has demonstrated that M2-branes
that wrap the intersection of the `holomorphic' surface
$F(e^{-it}z_1, e^{-it}z_2)=0$ with the unit 5 sphere are all
$SU(4/1)$ invariant. Similarly, supersymmetric giant gravitons in
$AdS_4\times S^7$ are M5 branes that sit at the centre of $AdS_4$.
Let $S^7$ be regarded as the unit sphere in $R^8$. $H_i$ ($i= 1
\ldots 4$) are rotations in mutually orthogonal planes in $R^8$. A
simultaneous rotation by ${\pi \over 2}$ in each of these directions
is a complex structure on this space. Mikhailov has demonstrated
that M2-branes that wrap the intersection of the `holomorphic'
surface $F(e^{-it}z_1, e^{-it}z_2, e^{-it}z_3, e^{-it} z_4)=0$ with
the unit 5 sphere are all $SU(4/1)$ invariant.

We now turn to the quantization of giant gravitons. Following \our\
we first regulate the space of holomorphic functions on $C^2$ or
$C^4$. Let $P^{2}_d$ or $P^4_d$ refer to the set of holomorphic
polynomials in $C^{2}$ or $C^4$  whose degree less than or equal to
$d$. $P_d$ is a linear vector space of dimensionality $n_d ={d+2
\choose 2}$ or ${d+4 \choose 4}$. In this section we demonstrate
that as a symplectic manifold, $P^{2}_1$ and $P^4_1$ are
respectively the spaces $CP^{2}$ or  $CP^{4}$ with symplectic form
(derived by restricting the dynamical symplectic form from the Born
Infeld and Wess Zumino action to our subspace of solutions)
$N\omega_{FS}$ where $\omega_{FS}$ is the Fubini Study form on
$CP^2$ or $CP^4$. It follows from this result, together with the
formal arguments of sections 2-4 in \our, (all of which apply to the
context of this note) that $P_d$ is the symplectic manifold
$CP^{n_d-1}$ with a current (distributional) symplectic form in the
cohomology class of $N \omega_{FS}$ (see \our). In particular the
partition function over the Hilbert space obtained from the
quantization of this symplectic manifold is (see section 4 of \our)
the power of $p^N$ in \eqn\pfmfreg{ \prod_{n_1, n_2=0}^{n_1+n_2\leq
d} {1 \over 1-p e^{-n_1 \mu_1 -n_2 \mu_2}}} where
\eqn\pfdef{Z_{5}^N=\Tr e^{-\mu_1 H_1-\mu_2 H_2}} for the case of the
M5 brane, and in \eqn\pfmtreg{ \prod_{n_1, n^2, n_3,
n_4=0}^{n_1+n_2+n_3+n_4 \leq d} {1 \over 1-p e^{-n_1 \mu_1 -n_2
\mu_2 -n_3 \mu_3 -n_4 \mu_4}}} where \eqn\pfdeft{Z_{2}^N=\Tr
e^{-\mu_1 H_1-\mu_2 H_2 -\mu_3 H_3 -\mu_4 H_4}.} for the case of the
M2 brane.

\pfmfreg\ and \pfmt\ are the partition functions obtained from the
quantization of regulated classes of Mikhailov's solutions. In order
to remove the regulator we simply take the limit $d \to \infty$;
under this limit \pfmfreg\ and \pfmtreg\ reduce to our proposals
\pfmf\ and \pfmt\ for the partition functions over appropriate
states in the M5 and M2 brane theories respectively.

In the rest of this section we will construct the manifold $P^{2}_1$
and $P^{4}_1$ of `linear' Mikhailov solutions.

\subsec{Quantization of Linear Polynomials}

In this section we construct and quantize the intersections of
\eqn\linpol{\sum_{i=1}^{k+1} e^{-i {t \over R}} c_i {y_i \over R} -1
= 0} in $C^k$ (here $y_i$s are the coordinates on $C^k$) with
\item{a)} with the sphere of radius $R$ in $C^{k+1}$  and
\item{b)} with the sphere of radius $R$ in $R^{2k+3}.$
We follow the procedure of  \our; see also \refs{\ggsumit,
\ggouyang} for related work.

We will quantize our manifold of solutions with respect to the
canonical symplectic form  obtained from the action \actnb\ (recall
that \actnb\ describes the motions of probe M5 branes in $AdS_4
\times S^7$, probe M2 branes in $AdS_7\times S^4$ and probe D3
branes in $AdS_5 \times S^5$ for appropriate values of the
parameters $m$ and $n$ defined before \actnb; the sphere referred to
above may be identified with the sphere in \scaledmet).

The action \actnb\ may be written out in more detail as \eqn\ga
{{\cal S} = {N\over {R^{n+1} \Omega_n}} \int \sqrt{-g}\ d^n\sigma\
dt + \int_{S^{n+2}} d^n \sigma A_{\m_0 \m_1 \cdots \m_n}
\dot{x^{\m_0}}{{\partial x^{\m_1}}\over {\partial \sigma _1}} \cdots
{{\partial x^{\m_n}}\over {\partial \sigma _n}}} Here $x^{\m}$s are
coordinates on $S^{n+2}$ and $\sigma$s are coordinates on the world
volume of the brane $\Sigma^n$. $g_{ij}$ is the induced metric on
$\Sigma^n$. $A$ is the $(n+1)$ form such that $dA = F = {{2\pi
N}\over{R^{n+2}\Omega_{n+2}}}\epsilon$ where $\epsilon$ is the
volume form on the $S^{n+2}$. We will find it convenient to work
with the rescaled complex variables $z_i={y_i \over R}$.  By a
$U(k+1)$ rotation any linear polynomial of the form \linpol\ can be
expressed as \eqn\linrot{c_0 e^{-i {t \over R}}z_{k+1} -1 = 0} where
$|c_0|^2 = \sum_i|c_i|^2$. The intersection of \linrot\ with an
$S^{n+2}$ of radius $R$, centered about the origin,  is an $S^n$ of
radius $R\sqrt{1-{1\over{|c_0|^2}}}$. It is not difficult to check
that \linrot\ satisfies the equations of motion that follow from
\ga. Further, the energy (momentum conjugate to $-t$) and the
angular momenta on this solution are (see Appendix A)
\eqn\enmom{\eqalign{E =& {N\over R}\left(1-{1\over
|c_0|^2}\right)^{{n-1}\over 2} \cr
                    P_{\theta} =& N\left(1-{1\over |c_0|^2}\right)^{{n-1}\over 2} \cr
                    E =& {1\over R}P_{\theta}}}

Let us now return to the general linear solution (the solution
parameterized by arbitrary complex coeficeints $c_i$). The
symplectic form is a closed two-form on the parameter space formed
by the coefficients $(c_i)$ of the solution and the symplectic form
must respect the $U(k+1)$ invariance of the action. These conditions
constrain the symplectic form to be of the form \eqn\symfrm{\omega =
f(|c|^2)\ \left({{d\bar{c^i} \wedge dc_i}\over {2i}}\right) +
{f^{\prime}}(|c|^2)\  \bar{c^i}c_j\ \left({{d\bar{c^j} \wedge
dc_i}\over {2i}}\right)} This function $f$ can be expressed as a sum
of two terms one coming from the Born-Infeld part of the action
(denoted by $f_{BI}$) and one coming from the Wess-Zumino term
(denoted by $f_{WZ}$). Below (in most of the rest of this section)
we compute $f_{BI}$ and $f_{WZ}$ following \our. We find
\eqn\finf{\eqalign{f_{BI}(x) =& -2N \ \ {1\over x^2}\
\left(1-{1\over x}\right)^{{{n-1}\over 2}} \cr
                   f_{WZ}(x) =&  -2N \ {{\left({1-{1\over x}}\right)^{{n+1}\over 2}}\over {x}}\cr
                   f(x) =& f_{BI}(x) + f_{WZ}(x)\cr
                        =& {-2N \over x}
                        \ {\left({1-{1\over x}}\right)^{{n-1}\over 2}}\
                        }}
Now since $|c|^2f(|c|^2)$  monotonically increases from $0$ to $2N$
as $|c|^2$ varies from $1$ to $\infty$, it is possible to make a
(non holomorphic) change of variables to convert the symplectic form
$\omega$ into to $N$ times the standard Fubini Study form
$\omega_{FS}$. Explicitly \eqn\sympform{ \omega= -2 N  \left[ {1
\over 1+ |w|^2} \left({{d\bar{w^i} \wedge dw_i}\over {2i}}\right)
-{1 \over (1+|w|^2)^2} \  \bar{w^i}w_j\ \left({{d\bar{w^j} \wedge
dw_i}\over {2i}}\right) \right].} where  \eqn\defw{w^i= c^i
\sqrt{{{\tilde f}(|c|^2) \over 1-|c|^2 {\tilde f}(|c|^2)}}} and
${\tilde f}(x)=f(x)/2 N$.

Note that the `hole' $|c|<1$ has been contracted away to a point in
the `good' $w$ variables (see \our\ for a detailed discussion of the
same phenomenon in a different context). The quantization of our
space is now standard in $w$ variables. The Hilbert space is given
by the holomorphic polynomials of $(k+2)$ variables $(w_i$ and $
 1)$ with $k$ charge operators $L_i = w_i\partial_{w_i}$. This is
identical to the Hilbert space of $N$ identical non-interacting
bosons whose single particle Hilbert space consists of $(k+1)$
states with the $i$th state having charge one under $L_i$ and charge
zero under all others and the $(k+2)$th state having all charges
zero.

In the rest of this section we present the computations that lead to
\finf.

\subsec{Calculation of $\omega_{BI}$} It is convenient to calculate
$\omega_{BI}$ in the rotated coordinate system where linear
polynomial takes the form of \linrot. As coordinates on the $(unit\
S^{n+2} = \Omega_{n+2})$ we will choose  \eqn\b { z_{k+1} = Z = \rho
e^{i\theta}} together with $[Z_1,Z_2,...,Z_{k},(H)]$ where  \eqn\c{
{[z_1,z_2,...,z_{k},(h)] \over \sqrt{1-\rho ^2}} \equiv
[Z_1,Z_2,...,Z_{k},(H)]. } Note that $H^2+ \sum_{i=1}^k |Z_i|^2=1$.
Of course the coordinate $H$ simply does not exist - and so may be
set to zero in all equations -  when $n$ is odd.

Since $\omega_{BI}$ is an exact form it can be expressed as
$d\Theta_{bI}$ where $\Theta_{BI}$ is an one-form given by
\eqn\thetabi {\Theta_{BI} = \int d^n\sigma (P_Z \delta Z +
P_{\bar{Z}} \delta \bar{Z})} where $Z = z_k$, $P_Z$ is the momentum
corresponding to the coordinate $Z$ evaluated at the solution given
by \linrot, $\delta Z$ and $\delta\bar Z$ is the fluctuation of $Z$
and $\bar Z$ (to the leading order) as the coefficients of the
linear polynimial are varied infinitesimally, and we have used the
fact that $P_{Z_i}= P_{\bar{Z}_i}= P_H=0$ on the solution \linrot.

$P_Z$ and $P_{\bar Z}$ can be computed from the $BI$ part of the
lagrangian which is given by
 \eqn\je {{\cal L}_{BI} =
  -{N\over{\Omega_n}}\ \sqrt{g^{\Omega_n}}\
  {{(1-Z{\bar Z})^{n\over2}}\over R}
  \sqrt{1-R^2\ {{{\bar{Z}}^2{\dot Z}^2 +Z^2{\dot{\bar Z}}^2
  +2(2-Z{\bar Z}){\dot Z} {\dot{\bar Z}}}\over{4(1-Z{\bar Z})}}}}
(\je\ may also be rewritten in terms of $\rho$ and $\theta$; see
Appendix A)

Therefore \eqn\l {P_Z = {{\partial {\cal L_{BI}}\over{\partial {\dot
Z}}}=
(i){N\over{\Omega_n}}\sqrt{g^{\Omega_n}}{(1-Z\bar{Z})^{{{n-3}\over
2}}}\left(Z {{{\bar{Z}}^2}\over 4} - {\bar Z} ({{2-Z\bar{Z}}\over
4})\right) }} where we have used ${\dot Z}={i Z \over R}$ on
\linrot. $P_{\bar{Z}}$ is simply the complex conjugate of $P_Z$.

We will find it useful to compute $\Theta_{BI}$ in terms of the
variables $c_i$. Consider the giant graviton \eqn\ggsf{ \left( c_0 +
\delta c_0 \right)  e^{i {t \over R}} z_{k+1} + \sum_{i=1}^k \delta
a_i e^{i{t \over R}} z^i=1.} This configuration is a small
fluctuation about \linrot; the corresponding variation in $Z$ is
given by \eqn\fluct{\delta Z = - {e^{i{t\over R}}\over{c_0
^2}}\delta c_0 -{{\sqrt{1-Z\bar Z}}\over {c_0}}\left(Z_1\delta a_1 +
Z_2 \delta a_2 + \ldots + Z_k \delta a_k \right );}  $\delta \bar Z$
is given by the complex conjugate expression.

 Using these expressions
\eqn\thetafi{\Theta _{BI} = N \left({i\over 2}\right) {1\over
{|c_0|^4}} \ \left(1-{1\over {|c_0|^2}}\right)^{{{n-1}\over 2}}\
\left(\bar{c_o} \delta c_0 - c_0 \delta \bar{c_0}\right)} (all terms
proportional to $\delta a_i$ evaluate to zero upon integrating over
the sphere).

Substituting $|c_0|^2 \rightarrow \bar{c^i}c_i = |c|^2$, $\bar{c_0}
\delta {c_0} \rightarrow \bar{c^i}dc_i$ and $c_0 \delta \bar{c_0}
\rightarrow c_i d{\bar{c^i}}$ \eqn\n {\Theta _{BI} = N {i\over 2}
{1\over {|c|^4}}\ \left(1-{1\over {|c|^2}}\right)^{{{n-1}\over 2}}\
\left(\bar{c^i} \delta c_i - c_i \delta \bar{c^i}\right)}

\eqn\o {\omega_{BI} = d\Theta _{BI} = f_{BI}(|c|^2)\
\left({{d\bar{c^i} \wedge dc_i}\over {2i}}\right) +
{f^{\prime}_{BI}}(|c|^2)\  \bar{c^i}c_j\  \left({{d\bar{c^j} \wedge
dc_i}\over {2i}}\right)}

Where \eqn\p {f_{BI}(x) = -2N \ {1\over x^2} \left(1-{1\over
x}\right)^{{{n-1}\over 2}}}

\subsec{Calculation of $\omega_{WZ}$} As argued in appendix C.1 of
\our\  $\omega_{WZ}$ can be written as ${2\pi
N}\over{R^{n+2}\Omega_{n+2}}$ times the volume swept out by the two
deformations of the brane surface. The volume of the giant graviton
$R^n\Omega_n(1-Z\bar{Z})^{n\over 2}$; the volume swept out when the
graviton is deformed is \eqn\ab {V_{deformed}=-\int d^n \sigma
R^{n+2}\sqrt{g^{\Omega_n}}\ (1-\rho ^2)^{{n-1}\over 2}\  \rho \delta
\rho \wedge \delta \theta  } Here $\rho$ and $\theta$ are the radial
and angular directions in unit $S^{n+2}$ that are perpendicular to
$S^n$. Using \b\ \eqn\bb {\omega_{WZ} = -{{2\pi N}\over
{R^{n+2}\Omega_{n+2}}}\ V_{deformed} = {{2\pi N}\over
{\Omega_{n+2}}}\int d^n \sigma \sqrt{g^{\Omega_n}}\
(1-Z\bar{Z})^{{n-1}\over 2}\ \left({{\delta \bar{Z} \wedge \delta
Z}\over {2i}}\right)} Using \fluct\ \eqn\cb {\omega_{WZ} = -{{2\pi
N}\over {\Omega_{n+2}}}\int d^n \sigma \sqrt{g^{\Omega_n}}\
(1-Z\bar{Z})^{{n-1}\over 2}\ \left({{\delta \bar{c_0} \wedge \delta
c_0}\over{ 2 i |c_0|^4}} + {{(1-Z\bar{Z})}\over {|c_0|^2}}\
\sum_{i=1}^{k-1} \left(|Z_i|^2 {\delta \bar{a_i} \wedge \delta a_i
\over 2 i}\right)\right)} Now
$$\int d^n \sigma \sqrt{g^{\Omega_n}} |Z_i|^2 = {2\over {n+1}}\
\Omega _n$$ .
 $$\delta \bar{c_0} \wedge \delta c_0 \rightarrow \bar{c^i}c_j\  \left({{d\bar{c^j} \wedge dc_i}\over {|c|^2}}\right)$$ and
$$\sum_{i=1}^{k-1} \left(\delta \bar{a_i} \wedge \delta a_i\right) \rightarrow d\bar{c^i} \wedge dc_i - \bar{c^i}c_j\  \left({{d\bar{c^j} \wedge dc_i}\over {|c|^2}}\right)$$
Using these equations we find \eqn\db {\omega_{WZ} = f_{WZ}(|c|^2)\
\left({{d\bar{c^i} \wedge dc_i}\over {2i}}\right) +
{f^{\prime}_{WZ}}(|c|^2)\  \bar{c^i}c_j\  \left({{d\bar{c^j} \wedge
dc_i}\over {2i}}\right)}
 Where
\eqn\eb {f_{WZ}(x) = -2\pi N\  \left({{2}\over{n+1}}\right)\
{{\Omega_n}\over {\Omega_{n+2}}}\  {{\left({1-{1\over
x}}\right)^{{n+1}\over 2}}\over {x}} = -2N \ {{\left({1-{1\over
x}}\right)^{{n+1}\over 2}}\over {x}}}

\newsec{Discussion}

In this note we have argued that the spectrum of $1/4^{th}$ BPS
states of the world volume theory and the ${1/8^{th}}$ BPS states of
the theory on M2 branes is very similar to the ${1/8}^{th}$ BPS
cohomology of ${\cal N}$=4 Yang Mills theory. One point of
difference is that the M5 and M2 brane supersymmetric spectra
consists entirely bosonic states, while the Yang Mills cohomology
includes fermions. The reason for this difference can be traced to
the fact that the Euclidean rotation group, $SO(d)$, is a reducible
in $d=4$ but irreducible in $d=6$ and $d=3$.  The BPS states studied
in this note  are singlets under $SO(6)$ or $SO(3)$ on the M5 or M2
worldvolume, and so are purely bosonic. $1/8^{th}$ BPS states in
Yang Mills theory are singlets only under one of the two $SU(2)$s in
$SO(4)$; a condition that allows fermions to contribute to this
cohomology.

The partition functions $Z^N_5$ and $Z^N_2$ have interesting
behavior in the large $N$ limit (see \KinneyEJ\ for a longer
discussion of the analogous behaviour of $Z_{3}^N$ in Yang Mills
theory). For simplicity let us set $\mu_1= \mu_2 =\nu$ in and
$\mu_1=\mu_2=\mu_3=\mu_4=\chi$ (in \pfmf\ and \pfmt) for discussion
of this paragraph. At fixed $\nu$ and $\chi$, \pfmf\ and \pfmt\
simply reduce to the relatively structureless formulas \gravf\ and
\gravt\ in the large $N$ limit. If, however, the large $N$ limit is
taken keeping ${\tilde{\nu}} ={\nu N^{{1 \over 2}}} $ or
${\tilde{\chi}} = \chi N^{{1 \over 4}}$, $Z^N_5$ and $Z^N_2$ undergo
sharp Bose Einstein type phase transitions at  ${\tilde \nu}$ and
${\tilde \chi}$ of order unity. In the large ${\tilde \nu}$ or
${\tilde \chi}$ `Bose condensed' phase we recover the partition
functions \gravf\ and \gravt. However the small ${\tilde \nu}$ or
${\tilde \chi}$ phase has different thermondynamics. It would be
interesting to understand the gravity dual of this new phase (see
\our\ for a related discussion).

The BPS states we have investigated in this note were easily
analyzed because they controlled by the relatively tame `diagonal'
dynamics of the Coulomb branch. 1/16 BPS states in these theories
are almost certain to explore the non Abelian dynamics of these
theories in a detailed way. While the investigation of these states
appears to be a rather difficult problem, it holds the promise of
substantial pay offs.

{\bf Acknowldegements}

We would like to thank J. Bhattacharya, N. Dorey, L. Grant,  S. Lahiri, 
S. Kim, G. Mandal and S. Trivedi and all the students in the TIFR theory room
for useful discussions. We also thank S. Lahiri and G. Mandal for comments 
on an advance version of this manuscript. The work of S.M. was supported
in part by a Swarnajayanti Fellowship. We would like to thank the
University of Hyderabad for hospitality while this paper was being
written. We must also acknowledge our debt to the steady and
generous support of the people of India for research in basic
sciences.

\appendix{A}{Charges of the Giant Graviton}
In this appendix we review the computation of the  charges of the
giant graviton \linrot\ (see for example \refs{\ggsus, \ggsumit,
\ggrob, \ggaki, \ggouyang.}). Let $z_{k+1}=\rho e^{i \theta}$. The
metric on time $\times $ a sphere of radius $R$ may be written as
\eqn\ge {ds^2 = -dt^2 + R^2[{{d\rho ^2}\over{1-\r^2}} + \r^2
d\theta^2 + (1-\r^2)(d\Omega_n )^2]} where $(d\Omega_n )^2$ is the
metric on the unit $S^{n}$. In this appendix we study the dynamics
of a giant graviton that (at a given time) wraps the $S^n$ but is
located at some point in $\rho$ and $\theta$. The action for such a
giant graviton is given by \eqn\gf {{\cal S} = \int dt L = N\ [-
{{(1-\r^2)^{n\over2}}\over R} \sqrt{1-R^2({{{\dot \r}^2}\over
{1-\r^2}} + \r^2{\dot\theta}^2}) + (1-\r^2)^{{n+1}\over
2}\dot{\theta}]} ${\dot \theta}$ is constant on all solutions to the
equations of motion that follow from \gf. $\r=constant$ is a
solution to the equation of motion provided
 \eqn\gg{ \dot{\theta} = {1\over R}}
The momentum conjugate to $\theta$ on  this solution is given by
\eqn\gh{ P_{\theta} = {{\partial L}\over{\partial
\dot{\theta}}}|_{\dot{\theta} = {1\over R}} = N(1-\r^2)^{{n-1}\over
2.}} The energy of the solution is given by \eqn\gi { E =
\dot{\theta}P_{\theta} - L = {N\over R}(1-\r^2)^{{n-1}\over 2}}
Therefore for this solution obeys the `BPS' relation \eqn\bpst{{
{\rm Energy} = {1\over R}( {\rm angular\  momentum })}} For the case
of $AdS^7 \times S^4$ and $AdS^4 \times S^7$ the value of $R$ is
$\half$ and $2$ respectively. \bpst\ is literally the BPS bound for
these cases.

\listrefs
\end